\documentclass[preprint,authoryear,12pt]{elsarticle}

\usepackage{graphics}
\usepackage{graphicx}
\usepackage{epsfig}

\usepackage{amssymb}
\usepackage{amsthm}
\usepackage{amsmath}
\usepackage{lineno}
\usepackage{float}
\usepackage{textcomp}
\usepackage{verbatim}
\usepackage{mathtools}
\usepackage{natbib}
\usepackage[figuresright]{rotating}
\usepackage{tabu}
\usepackage{nomencl}
\usepackage{pifont}
\usepackage{geometry}
\usepackage{txfonts}
\usepackage{hyperref}
\usepackage{multirow}
\usepackage{chngcntr}
\usepackage{hyperref}

\newcommand{\ang}{\AA ngstr\"om}

\begin{document}

\begin{frontmatter}



\title{A new correlation between solar energy radiation and some atmospheric parameters}

\author[a]{Antonio Dumas}
\author[a]{Andrea Andrisani\corref{cor1}}
\author[a]{Maurizio Bonnici} 
\author[a]{Mauro Madonia}
\author[a]{Michele Trancossi}

\address[a]{University of Modena and Reggio Emilia, via Amendola n.2, Reggio Emilia 42122, Italy}

\begin{abstract}
The energy balance for an atmospheric layer near the soil is evaluated. By integrating it over the whole day period a linear relationship between the global daily solar radiation incident on a horizontal surface and the product of the sunshine hours at clear sky with the maximum temperature variation in the day is achieved. 
The results show a comparable accuracy with some well recognized solar energy models such as the \ang-Prescott one, at least for Mediterranean climatic area. Validation of the result has been performed using old dataset which are almost contemporary and relative to the same sites with the ones used for comparison.

\end{abstract}
\begin{keyword}
Global solar energy; Temperature based models ; \ang-Prescott ; Hardgreaves-Samani ; Bristow-Campbel.



\end{keyword}
\cortext[cor1]{Corresponding author. Tel.: +39-0522-522.}

\end{frontmatter}




\section{Introduction}
\label{sec:intro}
A detailed analysis and evaluation of the amount of solar energy incoming on the Earth surface by means of electromagnetic radiations is nowadays of primary interest. Present demands of food and renewable energy reclaim an intensive agricultural production and the implementation of solar energy systems more and more efficient. The global daily solar radiation on horizontal surface at a given location -- in the following $G$ -- is the most critical input parameter concerning crop growth models, evapotranspiration estimates and the design and performance of solar energy devices as well \citep{Doorenbos,Porter,Zekai}. So various efforts have been taken in order to achieve a detailed global map of daily solar radiation energy.\\ 
Actually, this task cannot be considered fully absolvable at present. The intensity’s flux of energy coming from the Sun at the Earth's distance\footnote{Actually, at one Astronomical Unit, which corresponds to the mean distance between Sun and Earth.}, the so called Solar Constant $G_{sc}$, is known to be in mean $G_{sc}=82 KJ/m^2/ min$ by satellite observations \citep{Willson}. With some simple calculations \citep{Allen} we can to take into account of the various incident angles of the solar rays on the ground and of the periodical variation of the Sun-Earth distance in order to calculate $G_0$, the extraterrestrial daily solar radiation on a horizontal surface (\ref{eq:G0}). However, due to interaction of solar radiation with the terrestrial atmosphere, this amount of energy is significantly far from being available at the Earth surface, and it is subject to continuous fluctuations of various nature.\\ 
If a ``pure'' astronomical computation for $G$ is not suitable, on the other end a set of direct measurements of the daily solar radiation performed on the ground at different locations, covering the Earth with sufficiency density, is not available at moment. The costs of maintenance for the relative instrumental apparatus, together with the great operational difficulty in keeping their calibration, makes the number of stations that can efficiently and accurately perform this task very low. In 1999, the ratio of weather stations collecting solar radiation data relative to those collecting temperature data were calculated to be 1:100 in United States and at global scale it was predicted to be 1:500, as cited by \cite{Thornton}.\\ 
In absence of direct measurements, $G$ is usually estimated by means of different methodologies. Some methods calculate $G$ from $G_0$ by means of physical models describing the various absorption and scattering phenomena occurring in the atmosphere, so determining its total transmittance \citep{Klein,Braslau,Hoyt}. Most of the recent works in this field concern the estimation of the aerosol’s contribution on it \citep{Gueymard,Gueymard2}. Others \citep{Cano,Eissa} evaluate the total transmittance by geostationary satellite images data, making use of statistical formulas between atmospheric transmittance, surface albedo and an index of cloud coverage over a specific location; with these methods a resolution of $5-7 km^2$ in constructing solar radiation maps can be achieved \citep{Petrarca}. Satellite data are also used in combination with atmospheric physical models in order to estimate $G$ \citep{Cogliani,Cogliani2}. For a review of various methods which make use of satellite data we suggest \cite{Pinker}. Other approaches apply neural networks as well, in order to treat various input atmospheric parameters in non-linear models \citep{Rehman,Benghanem} or to consider a minimal number of local parameters together with ancillary data from other similar sites \citep{Marti}. Stochastic analysis is often employed too \citep{Richardson,Srikanthan,Bechini}.\\     
Besides these methods, simple empirical formulas relating solar radiation with some meteorological parameters are widely used for their conceptual and computational simplicity as well as for their high efficiency and accuracy. With time a wide range of possible relationships of this kind has been proposed (see \citealp{Ahmad}, for a partial list). The most known formula is the \ang-Prescott formula (A-P in the following), which relates the total transmittance of the atmosphere over a given location in terms of the effective hours of sunshine by means of a linear expression \citep{Angstrom,Prescott}. Strictly connected with the A-P formula (\ref{eq:N0}) is the Albrecht formula (\ref{eq:albrecht}), as reported by \citep{Guerrini}.\\ 
A-P is globally recognize and its application is recommended in both the FAO paper 24 \citep{Doorenbos} and FAO-56 PM \citep{Allen} to estimate $G$ if the measured one is unavailable or of questionable integrity; despite of this, different modifications are been attempted in order to improve the efficiency of this relation. For example various studies \citep{Benson,Almorox,Almorox2,Liu} have considered and analyzed a time dependence, to a monthly or to a seasonal  scale, for its coefficients. Others have related the A-P coefficients to geographical or meteorological elements \citep{Albrecht,Glover,Halouani}, or have used non-linear relationships in sunshine hours variable \citep{Barbaro,Akinoglu}, or have introduced other meteorological parameters \citep{Swartman,Ojo,Sabbagh,Garg,Odobo}. In these latter cases, variables used from time to time are humidity, average or maximum air temperature daily, fraction of covered sky, daily precipitation. Although some authors claimed that these modified models performed better than the conventional A-P model, this was apparently not the case in many comparative studies \citep{Iziomon,Almorox,Yorukoglu}. Actually they assure a probably negligible gain in accuracy at a cost of losing the simplicity and convenience of the original model\footnote{This lack of significant improvements in adding more parameters to the \ang-Prescott relation is sometimes be attributed to the uncertainty inherent in the measurements of global solar radiation and sunshine hours \citep{Persaud} and to an inter-dependency among different meteorological variables or between these and the AP coefficients \citep{Liu2}.}.\\ 
Besides A-P equation, other empirical relations, connecting $G$ with daily temperature variations $\Delta T$  instead of sunshine duration, are widely used and recognized. Even if the sunshine-based methods are generally believed to be more accurate \citep{Iziomon,Trnka}, temperature-based methods are often preferred since temperature records are much more available than sunshine records. Among the various models present in literature, we mention the Hardgreaves-Samani model \citep{Hardgreaves,Samani} and the Bristow-Campbell model \citep{Bristow}, in the following H-S and B-P respectively. For their expression see (\ref{eq:hardgreaves}) and (\ref{eq:bristow}). These relations too have been subjected to various modifications which include adding further parameters or meteorological variables (we suggest \citealp{Liu3}, for a comparative study). Actually, almost all of the empirical relations between $G$ and $\Delta T$ used today derives from the Hardgreaves-Samani \citep{Hunt,Supit}, or from the Bristol-Campbell ones \citep{Donatelli}.\\
The equation we want to propose here, even if in the same class of those relating temperature variations with global solar radiation, is somewhat different from the H-S and the B-C, together with their derivations, since it is linear in $\Delta T$. Besides, it takes an important role the number of hours of extra-atmospheric insolation $N_0$, because our relation actually correlates $G$ with the product $N_0\cdot\Delta T$. This product defines a new atmospheric parameter $F$ called \emph{action of temperature variation}. This equation was first introduced by \citet{Dumas}, together with another linear relation between the sunshine duration and $F$.\\ 
The Dumas equation is almost contemporary with the more famous Hardgreaves-Samani and Bristow-Campbell ones; however it did not achieve the same consideration and diffusion, probably due to the fact that the original work was written in Italian. So we revise the analysis and the results obtained in the original article of Dumas, regarding data acquired from the Italian cities of Modena, Bolzano, Genoa, Naples, Venice, Trieste and Udine, to promote the variable $F$ hoping that this relation might be easily acquired by a larger number of researchers and users in different fields.\\ 
This article is structured as follows. By an analysis of energy balance in a system consisting of a layer of atmosphere and a surface layer of soil adjacent to it, a function is found between the solar radiation incident on the surface of soil and the daily temperature variation of the corresponding atmospheric layer. The most important results about linear regression performed on data from seven Italian cities are then presented, together with error analysis. A comparison with other studies, regarding the same sites and approximately the same years, in which the \ang-Prescott, the Albrecht and the Barbaro’s formula were applied, is also reported. In the conclusions we will summarize the results obtained in this work, and in the appendix we will mention some of the formula here discussed.






\section{Correlation equation}
\label{sec:2}
It's well known that the Earth is a ``cold body'' like all the planets, namely it does not produce energy but it disperses into the surrounding environment, the sidereal space, the energy coming from the Sun in form of electromagnetic radiation. Actually only part of the earth's surface and atmosphere interact with solar radiation in a generic instant of time, so that even if the overall system ``atmosphere-terrestrial surface'' has a steady thermo-energetic budget, its subsystems may have a periodic one instead. There are fluctuations with different periods and different intensity: daily cycles, seasonal, secular. We assume that the energy fluctuation has zero average into such suitably restricted subsystem, even if this is related to the cycle with a shorter period. This cycle can be located roughly in the period of rotation of the earth around its axis, at least for not very high latitudes.\\
We consider a subsystem constituted by an atmosphere layer, a thin layer of soil adjacent to it and a higher layer of atmosphere above to it. The first two layers have an extension and thickness such that
\begin{itemize}
	\item[a)] Sufficiently small thermal inertia is assumed and of the same order of magnitude for the layers.
	\item[b)] These layers can be considered flat and parallel.
	\item[c)] The temperature gradient is orthogonal to the layers.
	
\end{itemize}
Besides, we suppose that in the atmospheric layer no thermodynamics transformation involving phase changes or chemical reaction and no change in pressure are present.\\
Let $T_a$, $T_s$, $T_u$ the temperatures respectively of the air, of the soil and of the higher atmosphere layers, $H_a$, $H_s$ the flux of solar energy incident on the air and the soil layers, $\alpha_a$, $\alpha_s$ their solar radiation absorption coefficiets and $h$, $k$ the convention and conduction coefficients. With the above assumptions, the instantaneous energy balance in an atmospheric layer, per unit area, is
\begin{equation}
\label{eq:at-bal}
\frac{\partial \varepsilon_a}{\partial t}=\alpha_aH_u-h\left(T_a-T_u\right)-k\left(T_a-T_s\right)
\end{equation}
where the term on the left is the temporal variation of the energy content of the layer $\varepsilon_a$, while those on the right are respectively the solar energy absorbed and the heat exchanged per unit time by convection and conduction with higher atmosphere layers and with the soil surface. We neglect here the energy exchange by thermal radiation, since the air is almost transparent to infrared radiation, having wavelengths of about $5-10 \mu m$.\\
Likewise, for the soil layer we have:
\begin{equation}
\label{eq:sol-bal}
\frac{\partial \varepsilon_s}{\partial t}=\alpha_s H_s-k\left(T_s-T_a\right)+\dot q_c+\dot q_r
\end{equation}                           	                                              
where $\dot q_r$   and $\dot q_c$ are the amount of energy exchange per unit of time respectively with the deeper soil layers and for infrared radiation with the sky. These quantities are assumed to be constant and their sum will be henceforth indicated with $\dot q_d$.\\
From (\ref{eq:sol-bal}) we have
\begin{equation}
\label{eq:sol-bal2}
k\left(T_s-T_a\right)=\frac{\partial \varepsilon_s}{\partial t}-\alpha_s H_s-\dot q_d
\end{equation}
and inserting it into (\ref{eq:at-bal}), we obtain
\begin{equation}
\label{eq:at-bal2}
\frac{\partial \varepsilon_a}{\partial t}=\alpha_a H_a-h\left(T_a-T_u\right)-\frac{\partial \varepsilon_s}{\partial t}-\alpha_s H_s-\dot q_d
\end{equation}
The temporal trend of the energy content of each layer depends on their thermal inertia, since they are subjected to thermal stress at the same time interval. Therefore, for the hypothesis a, we can suppose that
\begin{equation}
\label{eq:at-sol}
\frac{\partial \varepsilon_s}{\partial t}=c+d\frac{\partial \varepsilon_a}{\partial t}
\end{equation}
We also assume, following the same reasoning,
\begin{equation}
\label{eq:at-sol2}
T_u=c'+d'T_a
\end{equation}
Besides, by some considerations about energy flux conservation we have that $H_s$ is a fraction of $H_a$, namely
\begin{equation}
\label{eq:frac}
H_s=\beta H_a
\end{equation}
with $\beta$ the total transmittance for the air layer. Observe that $\alpha_a$, $\alpha_s$ and $\beta$ takes into account of the multiple reflections to which solar rays are subjected inside the air layer. By inserting (\ref{eq:at-sol}), (\ref{eq:at-sol2}), (\ref{eq:frac}) in (\ref{eq:at-bal2}) and integrating over a period $p$ corresponding to a fluctuation cycle, we get
\begin{align}
\label{eq:sum1}
&0=\int_p\left(1+d\right)\frac{\partial \varepsilon_a}{\partial t}dt-\int_p\left(\alpha_s+\alpha_a/\beta\right)H_s dt+
\int_p \left(c-\dot q_d-hc'\right)dt+\int_p \left(1+d'\right)h T_a dt
\end{align}
that can be rewrite as
\begin{equation}
\label{eq:sum}
I_1+I_2+I_3+I_4=0
\end{equation}
where $I_i$ is the $i$th integral on the right of (\ref{eq:sum1}). For the evaluation of these integrals we have to choice a fluctuation's cycle for the system's energy under consideration: we have taken the shortest one, the daily cycle, for which $p=24h$.\\
Now the first term $I_1$  is zero for the remarks made earlier. By assuming that $\alpha_s$, $\alpha_a$ and $\beta$ vary indipendently from $H_s$ during the daylight and by taking their mean value, the second term is rewritable as
\begin{equation}
\label{eq:I2}
I_2=-\left(\alpha_s+\alpha_a/\beta\right)\int_p H_s dt=-\alpha_cG
\end{equation}
where $\alpha_c=\alpha_s+\alpha_a/\beta$ while $G$ is the global daily solar radiation incident on a horizontal plane.\\
About $I_3$, all the integrand terms are constant. Therefore, it can be assumed to be proportional by a constant to the period of fluctuation:
\begin{equation}
\label{eq:I3}
I_3=\int_P\left(c-\dot q_d-hc' \right)dt=gp
\end{equation}
As concerned the forth integral, the first obvious way to estimate it could be to consider the daily mean temperature again in order to obtain
\begin{equation}
\label{eq:I4}
I_4=h\left(1+d'\right)\int_p T_a dt=h_c p\bar T
\end{equation}
with $h_c=h\left(1+d'\right)$. Then by inserting (\ref{eq:I2}), (\ref{eq:I3}) and (\ref{eq:I4}) in (\ref{eq:sum}) we get
\begin{equation*}
-\alpha_c G+gp+h_c p\bar T=0
\end{equation*}
So we identify a relation between $G$ with respect to $\bar T$. It can be expressed as
\begin{equation}
\label{eq:ojo} 
G=a_1+b_1\bar T
\end{equation}
where $a_1$  and $b_1$ are empirical constants to be determined with local climate data. However, results carried out by applying eq. (\ref{eq:ojo}) actually show low correlation, so that they are not reported here. They confirms those obtained by \citet{Ojo} and, obviously, his conclusions.\\
So it would be better to estimate $I_4$ by means of a different parameter rather then the mean temperature. If we think of the trend of the air temperature $T_a$ as given by a thermal pulse that begins when the system is at its minimum temperature $T_m$, instead of a fluctuation around its mean value $\bar T$, then for $I_4$ we can write
\begin{equation*}
I_4=h_c \!\!\int_p\left(T_a-T_m\right)dt+ h_c\!\!\int_p T_m dt=h_c\!\!\int_p\left(T_a-T_m\right)dt+h_c T_m p
\end{equation*}
The first term on the right of the above equation gives the thermal variation due to the solar energy pulse, while the other is a sort of residual energy term. About the integral term, we can reasonably assume to be proportional to the amplitude of the pulse and to the time interval during which the system is stressed. For the amplitude it's natural to take the difference between the maximum and minimum temperature, $\Delta T=T_M-T_m$, registered during the period $p$, while we set the stress period equal to the number of hours of extra-atmospheric insolation $N_0$, as defined in (\ref{eq:N0}).\\ 
As concerned the term $h_c T_m p$, we can put it as a constant not varying with the cycles. Indeed $T_m$ gives the temperature of the air layer when no external strain -- the solar energy pulse -- is acting on the system. By hypothesis in a cycle the net energy variation of the layer is null, so that its temperature at the end of a cycle -- i.e. at the beginning of the next one as well -- is supposed to return to its original value\footnote{On the contrary, $T_M$ is the results of the intensity of the solar pulse, and it can vary from cycle to cycle.\\
About  $T_m$, it is well known that it actually assumes different values during different cycles, contrary to our assertion.  This is primarily due, from the various sources of fluctuations that affect this variable and that we have neglected in our hypothesis,  to the net daily energy variation, which  is null only in mean: it tends to increase from the spring equinox to the autumn equinox and to decrease during the rest of the year. One could takes account of this by assuming a time dependence for $a$ and $b$ in equation (\ref{eq:dumas}), at monthly or at seasonal scale. However, since we are interested in keeping these equations as simple as possible, and in consideration of what happens in the A-P relation, where an analogous time dependence for the coefficients takes no effective improvement in the prediction of global solar radiation \citep{Liu3}, we will omit such dependences in our analysis.} $T_m$, at least in the conditions assumed in the hypothesis. So we can put $h_c T_m p=cost.$  and we can write
\begin{equation*}
I_4=bN_0 \Delta T+cost.
\end{equation*}
with $b$ a proportionality constant. Equation  (\ref{eq:sum}) now reads
\begin{equation*}
G=a+bN_0 \Delta T
\end{equation*}
or
\begin{equation}
\label{eq:dumas}
G=a+bF
\end{equation}
having set $a$ as another constant and having defined $F=N_0 \Delta T$ the \emph{action of temperature variation}. We call this equation the \emph{first Dumas equation}.\\
With (\ref{eq:dumas}) we get a useful relation between the daily global solar radiation incident on a horizontal surface with the atmospheric variable $F$, given by the product of the daily temperature variation $\Delta T$ and the extra-atmospheric solar insolation $N_0$. Both of these parameters are very easily available. The first Dumas equation appears more appropriate than (\ref{eq:ojo}) since it correlates the intensity of an external action -- the solar energy pulse -- with the modifications that occur on the state of the system due to the same action. On the contrary the mean temperature in (\ref{eq:ojo}) takes into account the previous history of the system as well. The application of (\ref{eq:dumas}) to the data of some Italian sites -- Bolzano, Genoa, Naples, Venice, Trieste and Udine -- confirms that this relation is well suited for its tasks. The correlations and the MPEs (mean percentage errors) obtained with (\ref{eq:dumas}) are of comparable accuracy with those obtained by other authors with other methods for the same sites. They applied the A-P relation, the Albrecht's relation and the Barbaro's relation respectively, for periods of time comprehending the years 1971-1973 considered by us. The results are exposed in sections \ref{sec:4} and \ref{sec:5}. We refer to future works a comparison of (\ref{eq:dumas}) with other temperature-based models as the H-S and B-C. Some preliminary results can be found in \citet{Andrisani}.\\
Finally, since there is a correlation between daily solar energy incident and effective sunshine hours N -- e.g. A-P equation (\ref{eq:angstrom}) -- from (\ref{eq:dumas}) we deduce that there must be a linear correlation even between the effective sunshine hours $N$ and $F$
\begin{equation}
\label{eq:dumas2}
N=a_1+b_1 F
\end{equation}
with $a_1$ and $b_1$ other constants. We wil call this equation the \emph{second Dumas equation}. We tested it just for one station.
\section{Methodology in analysis of data}
\label{sec:3}                 
We have used the least squares methods to verify the goodness of the correlation and to determine the coefficients $a$, $b$, $a^{'}$ and $b^{'}$ in (\ref{eq:dumas}) and (\ref{eq:dumas2}). This methods is well known, and we refer to \citet{Rao} for a review. Here we just describe another useful variable about the variance.\\
Let $X$, $Y$ two variables. The regression line of $Y$ with respect to $X$ defines the new variable
\begin{equation*}
Y^{'}=a_X+b_XX
\end{equation*}
with $a_X$ and $b_X$ the empirical coefficients obtained by the least square analysis.  A new linear regression for the set of the calculated data $Y^{'}$ with respect to the observed ones $Y$ gives another variable
\begin{equation*}
Y^{''}=a_{Y}+b_{Y} Y
\end{equation*}
where $a_Y$ and $b_Y$ are new empirical coefficients.\\
Systematic errors are evidenced by the discrepancy of $a_Y$ and $b_Y-1$ from the null value. This discrepancy can be estimated by means of $Y^{''}-Y=a_Y+(b_Y-1)Y$. On the contrary, the discrepancy of $Y^{''}$ from $Y^{'}$ reveals the presence of random errors. So the systematic error percentage defined as
\begin{equation*}
S^2=\frac{\sum_i{\left(Y_i^{''}-Y_i \right)^2}}{\sum_i{\left(Y_i^{''}-Y_i^{'} \right)^2}+\sum_i{\left(Y_i^{''}-Y_i \right)^2}}\times 100
\end{equation*}
gives an indication in percentage for systematic bias contained within the model with respect to the total error.\\
The determination of the correlation has been made for the following groups of values for $G$ and $F$
\begin{enumerate}
\item daily values for each month;
\item	daily values averaged for ten days and for a month, for each year;
\item	daily values averaged for ten days and for a month, for all the years.
\end{enumerate}
The days affected by inconsistent or incomplete information were deleted from the analysis. The mean values were calculated and analyzed only when the available days  were more than half.\\
Actually, the number of days, ten or thirty, for which the mean values were performed was not constant, both for the reasons previous mentioned or because in order to match three decades for each month, the last decade may also be constituted by 8 or 11 days, depending on months.\\
For brevity, the results reported here are only those obtained from data relating to the last group, because they are constituted by a greater number of data, and then they are statistically more reliable.
\section{Analysis of Data}
\label{sec:4}
The experimental data regard seven meteorological stations, all placed in Italy: Modena, Bolzano, Genoa, Naples, Venice, Trieste and Udine. However, the weather station of Modena is not homogeneous with the other, both for the number of years then for the instrumentation used therein. For this reason its results are presented separately.
\begin{table}[ht]\small
\caption{Important parameters for Modena correlation data.}
\label{tab:cormo}
\begin{tabular*}{\hsize}{@{\extracolsep{\fill}}cccccccccc@{}}
& & & & & & & & &\\
 & $n$ & $R$ & $E(R)$ & $a$ & $E(a)$ & $b$ & $E(b)$ & $\sigma/G$ (\%) & $S^2$ (\%) \\
\hline
\multirow{3}*{Global energy} & 1 & 0.899 & 0.004 & -13.8 & 3.2 & 0.56 & 0.008 & 28.9 & 20.0\\
 & 10 & 0.963 & 0.005 & -19.7 & 2.5 & 0.73 & 0.013 & 16.0 & 8.2\\
 & 30 & 0.972 & 0.006 & -20.9 & 3.8 & 0.74 & 0.021 & 12.5 & 6.5\\ 
 & & & & & & & & &\\

 & $n$ & $R$ & $E(R)$ & $a_1$ & $E(a_1)$ & $b_1$ & $E(b_1)$ & $\sigma/N$ (\%) & $S^2$ (\%) \\
\hline
\multirow{3}*{Sunshine hours} & 1 & 0.804 & 0.007 & -83 & 0.1 & 0.037 & $5\cdot10^{-3}$ & 44.0 & 35.4\\
 & 10 & 0.925 & 0.009 & -30 & 0.17 & 0.034 & $9\cdot10^{-3}$ & 20.6 & 15.4\\
 & 30 & 0.960 & 0.009 & -11 & 0.02 & 0.033 & 0.001 & 8 & 13\\

\end{tabular*}
\end{table}

\begin{figure}[ht]\vspace*{4pt}

\centerline{\includegraphics[width=7cm]{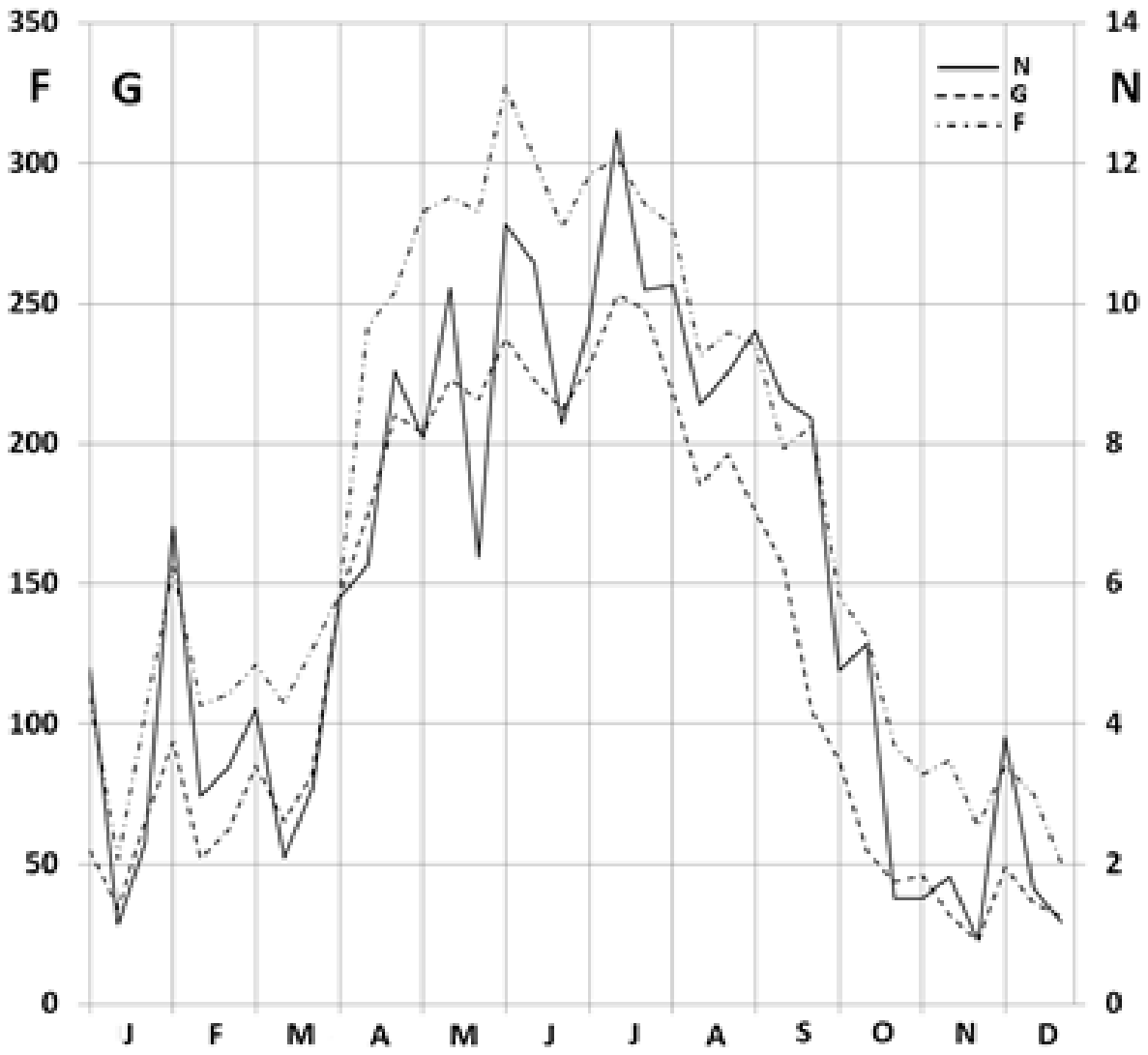}\hspace*{5mm}\includegraphics[width=7cm]{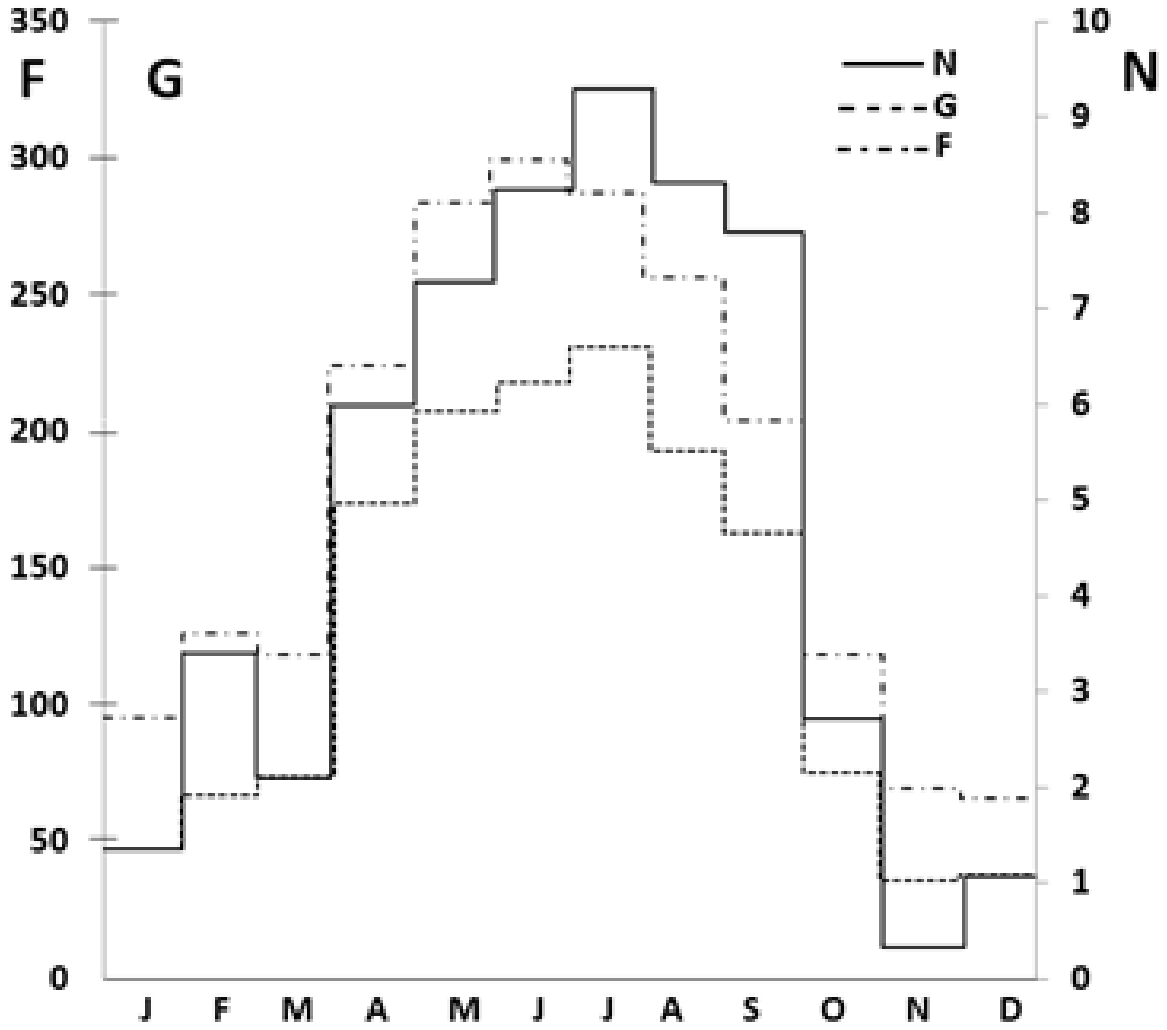}}
\caption{Ten days (right) and monthly (left) averaged values for $F$, $G$ and $N$ - Modena 1964 ($F=K∙h$, $G=kJ/m^2/day$, $N=h$).}
\label{fi:FGN_modena}
\vspace*{-6pt}
\end{figure}


\begin{figure}[ht]\vspace*{4pt}

\centerline{\includegraphics[width=7cm]{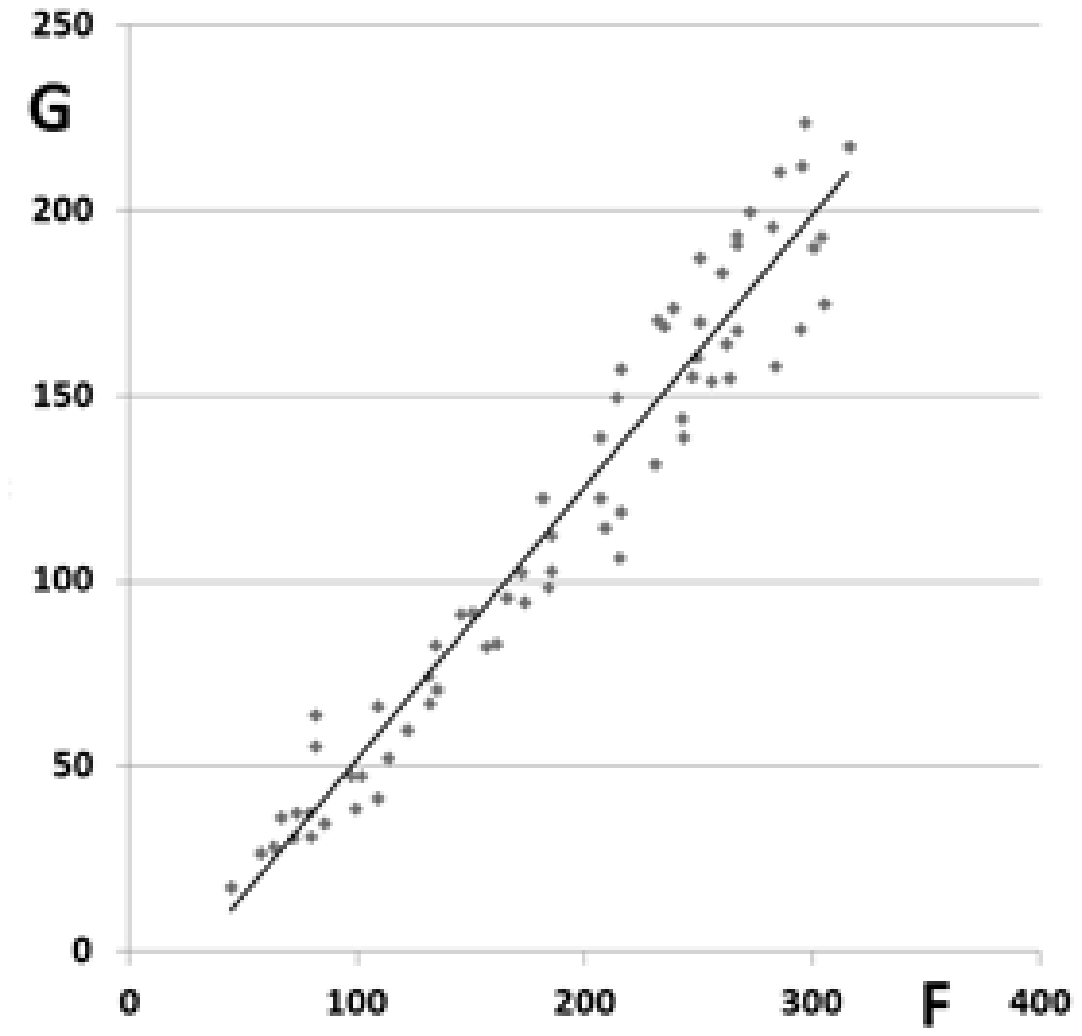}\hspace*{5mm}\includegraphics[width=7cm]{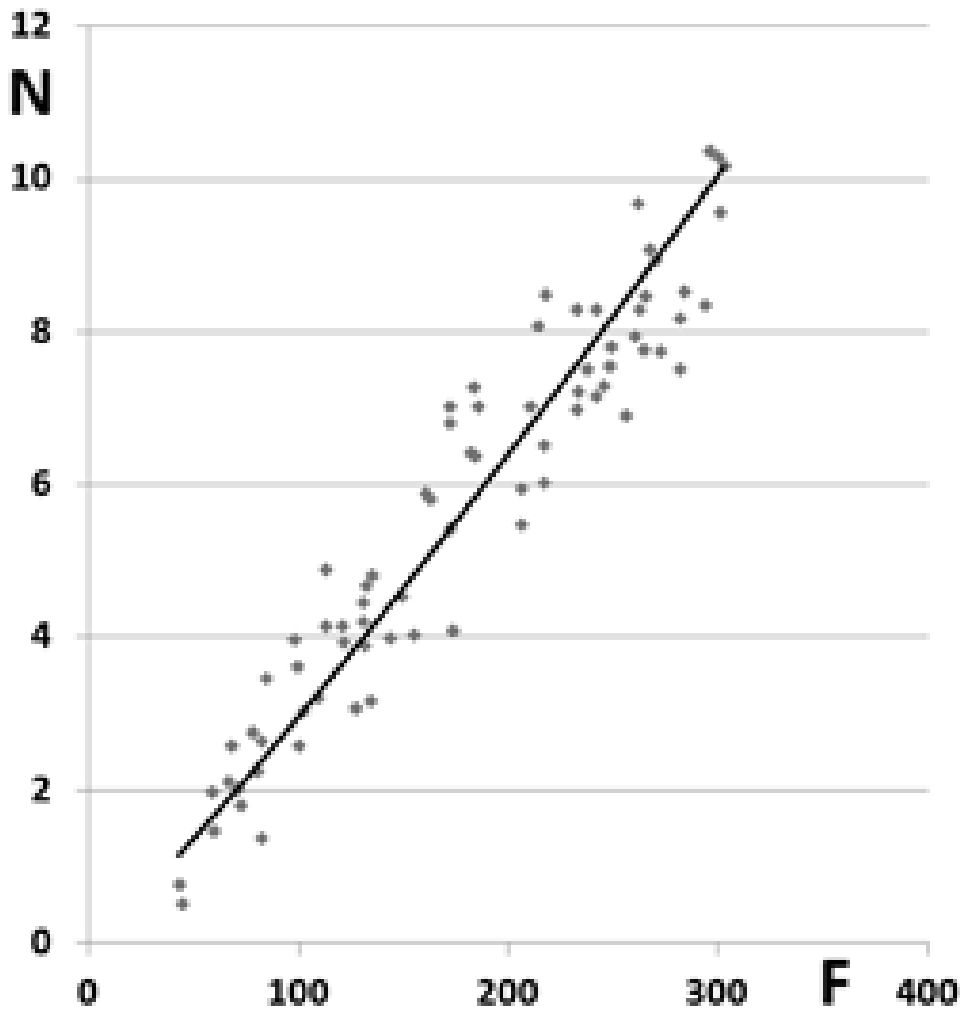}}
\caption{$G$ vs. $F$ (left) and $N$ vs. $F$ (right), monthly averaged values - Modena.}
\label{fi:GNvsFmodena}
\vspace*{-6pt}
\end{figure}

\begin{figure}[ht]\vspace*{4pt}

\centerline{\includegraphics[width=10.9cm]{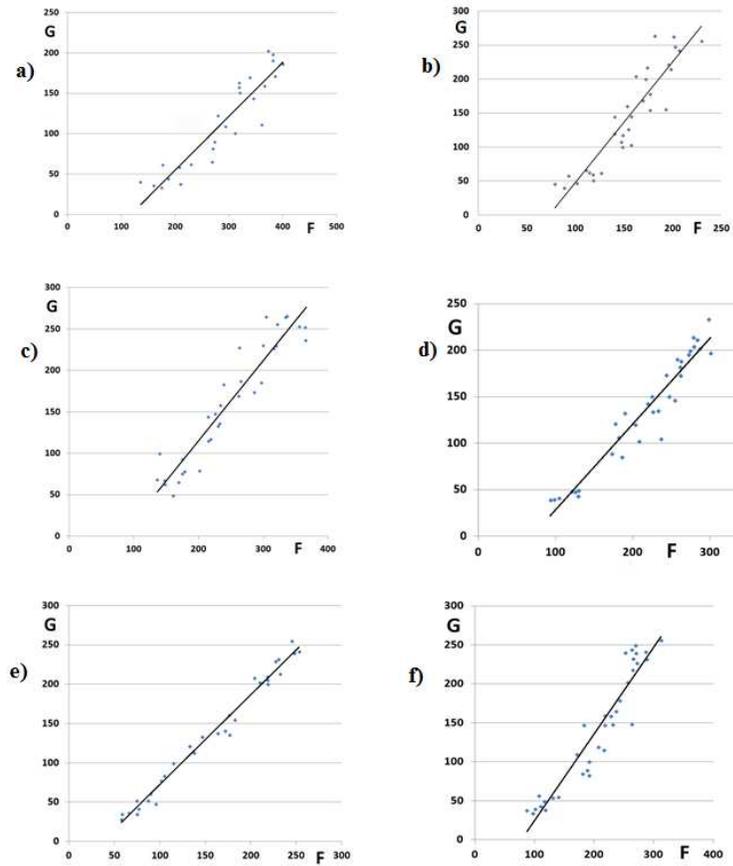}}
\caption{$G$ vs. $F$ for a) Bolzano, b) Genoa, c) Naples, d) Trieste, e) Udine, f) Venice. Data are monthly averaged.}
\label{fi:GvsF}
\vspace*{-6pt}
\end{figure}

\begin{table}[ht]

\caption{Important parameters about the correlation $G$-$F$ for the six meteorological stations considered.}
\label{tab:corcity}
\begin{tabular*}{\hsize}{@{\extracolsep{\fill}}lcccccccc@{}}
 & & & & & & & &\\
& $n$ & $R$ & $E(R)$ & $a$ & $E(a)$ & $b$ & $E(b)$ & $\sigma/G$ (\%)\\
\hline
\multirow{3}*{Bolzano} & 1 & 0.870 & 0.007 & -32.4 & 3.1 & 0.53 & 0.008 & 39.6\\
 & 10 & 0.920 & 0.015 & -59.6 & 8.4 & 0.61 & 0.025 & 27\\
 & 30 & 0.951 & 0.017 & -80.3 & 12.1 & 0.69 & 0.038 & 20.5\\ 
 & & & & & & & &\\
\multirow{3}*{Genoa} & 1 & 0.748 & 0.014 & -30.1 & 5 & 1.12 & 0.029 & 15.5\\
 & 10 & 0.885 & 0.022 & -128.4 & 13 & 1.59 & 0.084 & 29.3\\
 & 30 & 0.919 & 0.028 & -101.7 & 20.9 & 1.77 & 0.134 & 34.7\\
 & & & & & & & &\\
\multirow{3}*{Naples} & 1 & 0.805 & 0.011 & -18.8 & 4.2 & 0.73 & 0.017 & 50.6\\
 & 10 & 0.926 & 0.014 & -62.3 & 9.2 & 0.91 & 0.038 & 28.5\\
 & 30 & 0.954 & 0.016 & -76.1 & 13 & 0.97 & 0.05 & 22.6\\
 & & & & & & & &\\
\multirow{3}*{Trieste} & 1 & 0.873 & 0.007 & -32.6 & 3.1 & 0.69 & 0.013 & 38.5\\
 & 10 & 0.946 & 0.01 & -53.1 & 6.7 & 0.78 & 0.025 & 20.5\\
 & 30 & 0.966 & 0.01 & -58.6 & 9.5 & 0.8 & 0.038 & 16.3\\
 & & & & & & & &\\
\multirow{3}*{Udine} & 1 & 0.888 & 0.007 & -20.5 & 2.9 & 1.02 & 0.017 & 41\\
 & 10 & 0.977 & 0.005 & -38.5 & 4.2 & 1.14 & 0.025 & 16.7\\
 & 30 & 0.992 & 0.003 & -41 & 4.2 & 1.15 & 0.025 & 9.2\\
 & & & & & & & &\\
\multirow{3}*{Venice} & 1 & 0.849 & 0.008 & -36.8 & 3.6 & 0.86 & 0.017 & 47.3\\
 & 10 & 0.925 & 0.01 & -67.9 & 8.8 & 1.02 & 0.041 & 29.7\\
 & 30 & 0.951 & 0.017 & -82.8 & 13 & 1.09 & 0.063 & 23.8\\
\end{tabular*}
\end{table}

\begin{table}[ht]

\caption{Comparison of $R$ and $E(b)$ (in percentage) for three types of correlation - monthly mean values.}
\label{tab:compgue}
\begin{tabular*}{\hsize}{@{\extracolsep{\fill}}lcccccc@{}}
 & & & & \\
 & \multicolumn{2}{c}{This work} & \multicolumn{2}{c}{(\ref{eq:albrecht}) (\citeauthor{Guerrini})} & \multicolumn{2}{c}{(\ref{eq:angstrom}) (\citeauthor{Guerrini})} \\
 & $R$ & $E(b)$ (\%) & $R$ & $E(b)$ (\%) & $R$ & $E(b)$ (\%)\\
\hline
Bolzano & 0.951 & 5.5 & 0.86 & 4.4 & 0.60 & 11.0\\
 & & & & \\
Genoa & 0.919 & 7.6 & 0.92 & 2.3 & 0.82 & 4.8\\
& & & & \\
Naples & 0.954 & 5.2 & 0.90 & 3.6 & 0.64 & 9.7\\
& & & & \\
Trieste & 0.966 & 4.7 & 0.89 & 3.9 & 0.79 & 6.7\\
& & & & \\
Udine & 0.992 & 2.2 & 0.93 & 2.9 & 0.85 & 5.1\\
& & & & \\
Venice & 0.951 & 5.6 & 0.94 & 2.9 & 0.85 & 4.9\\
\end{tabular*}
\end{table}

\begin{table}[ht]\small

\caption{Percentage deviation of the estimated monthly mean value of G from the measured one ($\bullet$ This work, $\circ$ \citeauthor{Barbaro})}
\label{tab:compbar}
\begin{tabular*}{\hsize}{lcccccccccccc@{}}
 & & & & &\\
 & \multicolumn{2}{c}{Bolzano} & \multicolumn{2}{c}{Genoa} & \multicolumn{2}{c}{Naples} & \multicolumn{2}{c}{Trieste} & \multicolumn{2}{c}{Udine} & \multicolumn{2}{c}{Venice}\\
& $\bullet$ & $\circ$ & $\bullet$ & $\circ$ & $\bullet$ & $\circ$ & $\bullet$ & $\circ$ & $\bullet$ & $\circ$ & $\bullet$ & $\circ$\\
\hline
Jan	& 28.6 & 0.0 & 8.4 & 16.8 & 9.6 & 9.2 & n.a. & -1.9 & 24 & -6.7 & 37.4 & -3.8\\
Feb & -6.0 & 1.83 & -23.1 & 15.7 & 14.2 & 15.3 & 6.9 & 4.4 & n.a. & -3.8 & -22.4 & 3.9\\
Mar & 0.37 & -9.0 & -18.0 & 3.5 & 1.9 & 13.1 & 4.2 & -8.9 & 0.9 & -15.2 & -7.1 & -6.6\\
Apr & 7.2 & -3.2 & n.a. & 4.9 & 2.6 & 13.0 & n.a. & -2.4 & 3.3 & 1.9 & 1.6 & 9.5\\
May & 2.3 & -3.8 & 9.4 & 1.9 & 4.9 & 7.7 & 0.9 & -1.7 & 5.7 & -11.1 & -1.3 & -7.9\\
Jun & 9.1 & -0.7 & 16.5 & 3.5 & 3.9 & 6.5 & 2.2 & 0.9 & 4.4 & -3.8 & 10.0 & -3.4\\
Jul & 1.3 & -1.7 & 7.9 & 3.0 & 7.9 & 8.0 & 0.8 & -1.3 & 5.9 & 7.0 & 5.9 & -5.5\\
Aug & 5.2 & -1.7 & n.a. & 5.3 & 1.6 & 11.7 & 0.5 & -3.8 & 0.8 & -5.1 & 5.1 & -5.0\\
Sep & 6.7 & -3.9 & -6.7 & 7.2 & -4.3 & 14.4 & -7.4 & -3.3 & -8.1 & -8.3	& -4.7 & 0\\
Oct	& 29.9 & -1.6 & -23.2	& 10.0 & -11.5	& 11.3 & 3.8	& -5.3 & -32.6 & -10.7 & -31.1	& -3.5\\
Nov & 28.0	& 1.0	& -25.9 &	13.7 & -27.9 & 13.8	& -15.7	& -5.7 &	n.a. & 6.0 & -9.8	& 1.0\\
Dec & 25.0 & 3.7 & -1.5 & 18.6 & -28.3 & 9.3 & -9.1 & -7.1 & -4.9 & 0.0 & 13.7 & 2.4\\
\end{tabular*}
\end{table}
\subsection{Station of Geophysical Observatory of Modena}
All data were provided by the Geophysical Observatory, University of Modena (Osservazioni Meteorologiche, n. 87, 98 and 100-108). Years considered are 11, and specifically 1964, 1967, 1969-1977.\\
The daily data of solar energy incident on the horizontal plane $G$, the maximum and minimum temperatures and the effective hours of sunlight $N$ were detected respectively by a Moll-Gorczynski pyranometer, a maximum and minimum thermometer and a Campbell- Stokes heliograph. The instrumental error of a Moll-Gorczynski pyranometer depends mainly on the variation of the calibration factor that varies from $+2$\% to $-10$\% during the day \citep{Robinson}, so the average error of the daily incident energy can be estimated around $5$\%. Temperature measurements were generally affected by an error not less than $0.1$\textdegree C. Correspondingly, the error percentage on the temperature difference $T_M-T_m$ can be $10$\%  in winter months and less than $1$\% in summer months. The error percentage of $F$ assumes the same values. The effective hours of sunlight $N$ were measured with a Campbell-Stokes heliograph. They are affected by an absolute error almost constant in time, but variable from $3$\% to $15$\%. Actually, this error can be significantly higher, due to the complex method of calculation for the daily assessment of such variable \citep{Robinson}. Just for the year 1964, we have shown the temporal distributions in Figure \ref{fi:FGN_modena}. These distributions refer to average values of ten days and of a month for $G$, $N$ and $F$. The dispersion graph with the regression line for the monthly average values of $G$ and $F$ and of $N$ and $F$ are shown in Figure \ref{fi:GNvsFmodena}. Table \ref{tab:cormo} gives the values of the most important parameters about the correlations of $G$ and $N$ with $F$.
Columns are read in the following order: the correlation coefficients, the coefficients of the regression lines joined by their corresponding errors, the relative percentage of the standard deviation $\sigma$ and finally the systematic error percentage $S^2$.\\
Data in the table and in the graphs above show that the trend of $F$ is not uniform with respect to $G$ or $N$ in July. This difference is verified for the average values of ten days, but also for the monthly ones. It is not verified for all the other years. $N$'s distributions are more scattered than those corresponding to $G$, probably due to a greater experimental error in the calculation of daily sunshine hours. Similarly to what verified by \citet{Dumas2} about the A-P equation, (\ref{eq:dumas}) and (\ref{eq:dumas2}) are almost independent on the type of data used, i.e. the coefficients $a$ and $b$ are nearly the same both with the daily values than with the mean daily values. Nevertheless we have observed an increase in the correlation coefficient, a decrease in the standard deviation percentage and in the systematic error percentage with the increase in the averaging period.\\
The trend of systematic error percentage, together with the apparent casual anomaly in July 1964, suggests a possible dependence of the daily solar energy by some other weather variable. A dependence on another variable means that the thermal balance is probably overly simplified. In every case, this dependence is low for the daily values, and it also tends to vanish  when the period of  the average values of temperature increases.
\subsection{The Air Force weather stations of Bolzano, Genoa, Naples, Udine, Trieste, Venice}
The study was extended to data related to weather stations in Bolzano, Genoa, Naples, Udine, Trieste and Venice for the years 1971-1973. As we has said, these data are also taken into consideration by other authors \citep{Barbaro,Guerrini}. The variable $G$ was now measured by a SIAP pyranometer. This instrument has a lower precision than the Moll-Gorczynski pyranometer, and it also needs more frequent calibration controls.\\ 
The correlations were obtained for the same types of grouped data; however, the results shown here are less extensive than in the previous case. The distribution of monthly mean values of daily incident radiation $G$ versus $F$ and the calculated regression line are compared for the six cities in Figure \ref{fi:GvsF}.\\
In Table \ref{tab:corcity} are shown the most significant parameters of correlations, both for daily values then for ten days and month mean values. The constancy of the correlation coefficients $a$ and $b$ results to be just a feature of Modena station. The correlation between $G$ and $F$ is extremely good for all stations, even if a comparison between Tables \ref{tab:cormo} and \ref{tab:corcity} emphasizes that the best correlation is obtained for Modena station. This is probably due to a greater accuracy in the experimental investigation and to a greater precision of the measurement equipment.\\
In Tables \ref{tab:compgue} and \ref{tab:compbar} a comparison between our results and those obtained respectively by \citet{Guerrini} and \citet{Barbaro} is reported. We have to mention that these comparisons have mostly an indicative value, since the number of years considered, the procedures used in the data rejections and the methods for the evaluation of correlation coefficients are in general different for the various authors.\\ 
In Table \ref{tab:compgue} below we show the correlation coefficients and the percentage error $E(b)$ of the slope $b$ of the regression line corresponding to eq. (\ref{eq:dumas}) in this paper, but corresponding to the A-P relationship (\ref{eq:angstrom}) and to the Albrecht equation (\ref{eq:albrecht}) both used in \citet{Guerrini}.\\
The comparison clearly shows that the A-P formula is the worst one; besides, the correlation coefficients for (\ref{eq:dumas}) are constantly higher than those of the Albrecht formula, while $E(b)$ is slightly higher for the majority of cases.\\
A further comparison was made with the results obtained by \citet{Barbaro} for the same stations (see Table \ref{tab:compbar}). They used the relation (\ref{eq:barbaro}). In this case, the only parameter for comparison is the $MPE$, the mean percentage error defined as
\begin{equation*}
MPE=\frac{G_{meas}-G_{calc}}{G_{meas}} \times 100
\end{equation*}
where with $G_{meas}$ and $G_{calc}$ we intend respectively the monthly mean measured values of $G$ and the monthly mean estimated values of $G$. As we can see, in the present work the $MPE$ tends to increase from summer months to winter ones, while in the case of the Barbaro et al. the $MPE$ value  is almost constant for all the months of the year, with a slight increasing trend in the intermediate  months.  So it seems that (\ref{eq:dumas}) is more accurate than (\ref{eq:barbaro}) in the middle period of the year, while the opposite is true for the colder months. Actually, a lesser performance for low values of ∆T, as it is usually registered during the coldest months, is a common characteristic of the temperature based models \citep{Liu3}.
\section{Conclusion}
\label{sec:5}
From an analysis of the results here reported about the estimations of $G$, it seems that a high degree of accuracy can actually be achieved with the Dumas relation. The data and graphs exposed above show that the correlation coefficient r is never below $0.8$ for daily data and never below $0.9$ for the less fluctuating mean monthly data. These last values were wide above the ones obtained with the Albrecht and the trustworthy A-P relation, even if they compare different variables. With respect to Barbaro's relation, the overall performance is the same, even if with different behavior in the cold and the warm months of the year. The other Dumas relation, correlating $N$ with $F$, showed an analogous good behavior for the site of Modena. For this site we have registered a little systematic error, for both (\ref{eq:dumas}) and (\ref{eq:dumas2}), up to $20$\% and $35.4$\% for the daily case, which decreases to $6.5$\% and $13$\% respectively for the monthly case. This is probably due to a limited influence of another atmospheric variable.\\
Actually, these first results must be confirmed with further studies, possibly regarding solar energy estimations for an increased number of sites, located in different climatic conditions, with modern experimental apparatus, and for a great number of years. In particular in these last case we could test and compare the prediction as well, together the description as in the present article, for the variable $G$ with the various models considered so far.\\ 
A comparison of (\ref{eq:dumas}) with the Hardgreaves-Samani and Bristow-Campbell models has to be considered too, since all these relations employ daily temperature variations $\Delta T$ in estimating $G$. Dumas equation is actually the simplest among these, relating linearly $G$ and $F$. Besides, from some preliminary studies \citep{Andrisani} it seems that the overall accuracy of the model is comparable with those obtained by the H-S and B-C models. In particular eq. (\ref{eq:dumas}) could perform better than the other based-temperature models on sites with low $\Delta T$ values. We will make a more detailed discussion about these suppositions in future works.
\appendix
\section{}
\label{sec:A}
We report here some of the formulas cited in the article. The extraterrestrial daily solar radiation on a horizontal surface $G_0$ is given from the Solar Constant $G_{sc}$ by
\begin{equation}
\label{eq:G0}
G_0=\frac{24\cdot 60}{\pi}\ G_{sc} d_r \left(\omega_s  \sin ⁡  {\phi}  \sin{\delta}+\cos{\phi}\cos⁡{\delta}\sin⁡{\omega_s}\right)
\end{equation}
with
\begin{align*}
&d_r=1+0.033 \cos⁡{\left(\frac{2\pi j}{365}\right)}\\
&\delta=0.409 \sin⁡{\left(\frac{2\pi j}{365}-1.39\right)}\\
&\omega_s=\arccos{\left(-\tan{\phi}\tan{\delta}\right)}
\end{align*}
\citep{Allen}. Here $\phi$ is the latitude of the location, $j$ is the number of the day in the year, and  $d_r$, $\delta$, $\omega_s$ are respectively the Sun-Earth distance, the solar declination and the sunset hour angle for that day. Observe that by $\omega_s$ we can estimate $N_0$, the astronomical day duration (in hours)
\begin{equation}
\label{eq:N0}
N_0=\frac{24}{\pi}\omega_s
\end{equation}
The expression for the \ang-Prescott formula\footnote{In the original paper \citet{Angstrom}, took for $G_0$ the total radiation under a real atmosphere in completely clear day, and set $a+b=1$, since for a perfectly clear sky ($N=N_0$) $G$ must be equal to $G_0$. For a theoretical derivation of the \ang-Prescott's formula see \citet{Njau}, while we suggest \citet{Martinez} for an historical review of this equation.} is the following
\begin{equation} 
\label{eq:angstrom}						
\frac{G}{G_0}=a+b\frac{N}{N_0}
\end{equation}
where $a$ and $b$ are dimensionless parameters depending on the location. They are usually given at priori or estimated by least square linear regression methods\footnote{Calibration of (\ref{eq:angstrom}) for a given location is usually done using monthly mean daily $G$ and sunshine hours; however sometimes daily data and ten days mean daily data are used as in the present work on Dumas relationships. A comparative study about various time-scale calibrations for the coefficients $a$  and $b$ can be found in \citet{Gueymard3,Liu3}.}, and their physical meaning consist in the total transmittance during a clear day ($N=N_0$) for $a+b$ and during a cover day ($N=0$) for $a$. The A-P coefficients have been extensively studied and calibrated worldwide; a topographic map of  these coefficients for Europe can be found at this link \url{http://supit.net/main.php?q=aXRlbV9pZD02OQ==}.\\
The Albrecht relation, as reported by \citet{Guerrini}, is very similar to the A-P relation. It is given by
\begin{equation}
\label{eq:albrecht}
G=a+b N
\end{equation}
with the obvious meaning of symbols.\\ 
Barbaro's relation \citep{Barbaro} makes use of a nonlinear expression between $G$ and $N$
\begin{equation}
\label{eq:barbaro}
G=K N^{1.24} h_n^{-0.19}+10550\left(⁡\sin{h_n}\right)^{2.1}+300\left(\sin{h_n}\right)^3⁡
\end{equation}
where $K$ is an empirical factor depending on the latitude of the location, while $h_n$ is the maximum height of the Sun on the $15$th of the month.\\
Finally we report the most known relationships which estimates the daily global solar radiation energy as functions of the  daily variation temperatures. The Hardgreaves-Samani \citep{Hardgreaves,Samani} relationship is given by
\begin{equation}
\label{eq:hardgreaves}
\frac{G}{G_0}=a+b \sqrt{\Delta T}
\end{equation}
while the Bristow-Campbell \citep{Bristow} relationship reads
\begin{equation}
\label{eq:bristow}
\frac{G}{G_0}=a\left[1-e^{-b \left(\Delta T\right)^{c}}\right]
\end{equation}

\bibliographystyle{model2-names}
\bibliography{bib-article}







\end{document}